\documentclass{emulateapj}

\newcommand{\lap}{$\lesssim$}

\newcommand{\za}{$z$$\approx$}
\newcommand{\zs}{$z$$\sim$}
\newcommand{\sqam}{arcmin$^2$}
\newcommand{\Ebv}{$E(B-V)$}
\newcommand{\um}{$\mu$m}
\newcommand{\s}{$\sim$}

\newcommand{\Rlim}{$\cal R$$_{lim}$}
\newcommand{\UGRI}{$U_n G {\cal R} I$}
\newcommand{\U}{$U_n$}
\newcommand{\G}{$G$}
\newcommand{\R}{$\cal R$}
\newcommand{\I}{$I$}

\newcommand{\UG}{$U_n-G$}

\newcommand{\GR}{$G-{\cal R}$}
\newcommand{\RI}{${\cal R}-I$}
\newcommand{\UGR}{$U_n G {\cal R}$}
\newcommand{\GRI}{$G {\cal R} I$}

\newcommand{\lbgs}{Lyman Break Galaxies}
\newcommand{\Lstar}{$L^*$}

\shorttitle{Keck Deep Fields. I.  Data}
\shortauthors{Sawicki \& Thompson}

\begin{document}


\title{Keck Deep Fields. I. 
Observations, Reductions, and the Selection of Faint Star-Forming
Galaxies at Redshifts \zs4, 3, and 2 \altaffilmark{1}}

\author{Marcin Sawicki\altaffilmark{2}} 
\affil{
Dominion Astrophysical Observatory, 
Herzberg Institute of Astrophysics,
National Research Council, 
5071~West Saanich Road, 
Victoria, B.C., V9E 2E7, 
Canada;
and
Caltech Optical Observatories, 
California Institute of Technology,
MS 320-47, Pasadena, CA 91125, 
USA
}

\email{sawicki@physics.ucsb.edu}

\author{
David Thompson} 
\affil{
Caltech Optical Observatories, 
California Institute of Technology,
MS 320-47, Pasadena, CA 91125, 
USA}
\email{djt@irastro.caltech.edu}

\slugcomment{Accepted for publication in ApJ}

\altaffiltext{1}{Based on data obtained at the 
W.M.\ Keck Observatory, which is operated as a scientific partnership
among the California Institute of Technology, the University of
California, and NASA, and was made possible by the generous financial
support of the W.M.\ Keck Foundation.}
\altaffiltext{2}{Present address: Department of Physics, 
University of California, Santa Barbara, CA 93106, USA}

\begin{abstract}
We introduce a very deep, \Rlim\s27, multicolor imaging survey of very
faint star-forming galaxies at \zs4, \zs3, \zs2.2, and
\zs1.7.  This survey, carried out on the Keck I telescope, uses the
very same \UGRI\ filter system that is employed by the Steidel team to
select galaxies at these redshifts, and thus allows us to construct
identically-selected, but much fainter, samples.  However, our survey
reaches \s1.5 mag deeper than the work of Steidel and his group,
letting us probe substantially below the characteristic luminosity
\Lstar and thus study the properties and redshift evolution of the
{\it faint} component of the high-$z$ galaxy population.  The survey
covers 169 \sqam\ in three spatially independent patches on the sky
and --- to \R$\leq$27 --- contains 427 \GRI-selected \zs4 \lbgs, 1481
\UGR-selected \zs3 \lbgs, 2417 \UGR-selected \zs2.2 star-forming 
galaxies, and 2043 \UGR-selected \zs1.7 star-forming galaxies.  In
this paper, the first in a series, we introduce the survey, describe
our observing and data reduction strategies, and outline the selection
of our \zs4, \zs3, \zs2.2, and \zs1.7 samples.

\end{abstract}


\section{INTRODUCTION}

By $z$\s1, when the Universe was about half its current age, many of
the properties of the present-day galaxy population were already in
place.  Although the rate of star formation in the Universe as a whole
was an order of magnitude higher at $z$$\sim$1 than it is today (Lilly
et al.\ 1996), there already existed a well-developed luminosity
function of quiescent galaxies (Lilly et al.\ 1995) as well as an
established population of galactic bulges and disks (Schade et al.\
1995, Brinchman et al.\ 1998) with normal-looking Tully-Fisher
rotation curves (Vogt et al.\ 1997).  Thus, while clearly much remains
to be learned about galaxies and galaxy evolution at $z$$<$1, we must
look to higher redshifts, $z$$>$1, to witness many of the key events
in the story of galaxy formation.

To efficiently reach beyond $z$$>$1 requires techniques that let us
robustly select high-$z$ galaxies, preferably with some --- even crude
--- redshift information, while minimizing contamination by the far
more numerous foreground objects.  Such selection is possible using
multi-color broadband photometry that is sensitive to the imprint on
galaxy spectra of coarse spectral features such as the Lyman break at
rest-912\AA, the Balmer and 4000\AA\ breaks at 3648--4000\AA, and ---
in the rest-frame infrared --- the H$^-$ opacity bump at 1.6\um\
(e.g., Sawicki 2002).  One such multicolor approach is the technique
of photometric redshifts in which the most likely redshift of an
object is estimated by comparing its observed and predicted broadband
spectral energy distributions (e.g., Connolly et al.\ 1995; Sawicki,
Lin, \& Yee 1997; Bolzonella et al.\ 2000; Sawicki 2002; Csabai et
al.\ 2003).  Another, even simpler, technique --- popularized and
shown to be very effective through extensive spectroscopic follow-up
by Steidel and collaborators (e.g., Steidel et al.\ 1996, 1999, 2003)
--- straightforwardly selects high-$z$ star-forming galaxies by their
distinctive colors in an optical color-color diagram.  Extensive
spectroscopic follow-up of such color-color selected Lyman Break
Galaxies (LBGs) has allowed Steidel and his collaborators to amass
very large samples of $\sim$10$^3$ {\it spectroscopically confirmed}
galaxies at $z$$\sim$3 (selection in the
\UG\ vs.\ \GR\ color space) and $z$$\sim$4 (selection in the \GR\ vs.\
\RI\ space) while a recent extension of the technique to lower
redshifts ($z$$\sim$1.7, $z$$\sim$2.2; Erb et al.\ 2003, Steidel et
al.\ 2004) has also started to yield promising results.

The imaging samples assembled by the Steidel team are designed for
efficient spectroscopic confirmation and, therefore, are limited to
relatively bright objects (\R\lap25.5 or \I\lap25).  Consequently,
they do not probe significantly fainter than the characteristic
luminosity, $L^*$, at which the galaxy luminosity function changes
slope.  

The transition at \Lstar\ in the galaxy luminosity function is likely
an imprint of how galaxies form and evolve and the comparison of
galaxies above and below \Lstar\ may tell us much about what different
processes are responsible for this evolution.  The mass function of
dark matter halos, predicted from simulations of dark-matter
clustering (e.g., Jenkins et al.\ 2001), is essentially a power law on
mass scales that encompass the range of galaxy masses; in contrast,
the observed luminosity function of galaxies, both at low redshift and
high, exhibits different behaviours above and below a characteristic
luminosity, \Lstar.  This different shape in the luminisity function
suggests that different mechanisms dominate the evolution of galaxies
above and below \Lstar and that therefore our understanding of galaxy
formation and evolution may profit from studying the evolution of not
just the bright end but also the faint component of the galaxy
population at high redshift.  Similarly, because the strength of the
clustering of dark matter halos depends on the halo mass, a careful
study of galaxy clustering properties as a function of both epoch {\it
and} galaxy luminosity may inform us about how the properties of dark
matter halos affect the star formation that occurs in the galaxies
that they host.

Motivated by the twin goals of studying the dependence of the galaxy
luminosity function and of galaxy clustering on epoch {\it and}
luminosity, we have carried out a very deep (\Rlim\s27), wide-area
(169 \sqam), multicolor (\UGRI) {\it imaging} survey of galaxies that
are significantly fainter than those reached in the well-known studies
by the Steidel team.  Our survey uses the {\it same} \UGRI\ filter
system that is used by Steidel and his collaborators for selecting
their \zs4, \zs3, \zs2.2., and \zs1.7 samples, but probes
significantly fainter --- up to 1.5 magnitudes, or a factor of 4 in
luminosity.  Spectroscopic follow-up of their color-color selected
samples has allowed Steidel et al.\ (1999, 2003, 2004) to precisely
characterize important quantities, including selection volumes and
fractions of low-$z$ interlopers. Our use of identical filters and
color-color selection permits us to apply this knowledge in a
relatively straightforward way to our samples. Thus, even without what
--- at the magnitudes of the objects in our survey --- would have been
{\it very} expensive spectroscopy, we can understand and correct for
the selection effects that are at play.

In the present paper --- which serves as the introduction to our
survey --- we describe our observations (\S~\ref{observations}) and
data reductions (\S~\ref{datareductions}), including an assessment of
the photometric completeness and surface brightness selection.  We
then outline the photometric selection of our \zs4, \zs3, \zs2.2, and
\zs1.7 galaxy samples (\S~\ref{LBGselection}) and compare and contrast
our survey with other deep imaging surveys
(\S~\ref{summarydiscussion}).  Subsequent papers in this series will
study the faint end of the high-$z$ galaxy luminosity function and the
clustering of faint galaxies at high redshift and will extend this
work into the near-IR.

Unless specifically stated otherwise, throughout this series of papers
we normalize fluxes on the AB magnitude system (Oke 1974) and adopt
$\Omega_M$=0.3, $\Omega_{\Lambda}$ = 0.7, and $H_0$=70
km~s$^{-1}$~Mpc$^{-1}$.

\section{SURVEY STRATEGY AND OBSERVATIONS}\label{observations}

The design of the survey reflects our goal of robustly studying the
evolution of the {\it faint} (sub-\Lstar) component of the
star-forming galaxy population at high redshift.  Briefly:
\begin{itemize}

\item {To reach {\it deep} into the sub-\Lstar\ population we carried out
very deep imaging using the LRIS imaging spectrograph on the 10m
Keck~I telescope.  }

\item {To {\it robustly} identify
high-$z$ star-forming galaxies and to ensure a smooth joining with the
Steidel et al.\ work at brighter magnitudes, we used the very same
filter set and selection techniques as are used in their larger, but
shallower, spectroscopically-calibrated surveys.}

\item {To avoid being dominated by small number statistics {\it and} by
cosmic variance the survey covers a large area (169
\sqam) that is split into three spatially-independent patches.  }

\end{itemize}
We call our survey the Keck Deep Fields (KDF).

\begin{figure}
\plotone{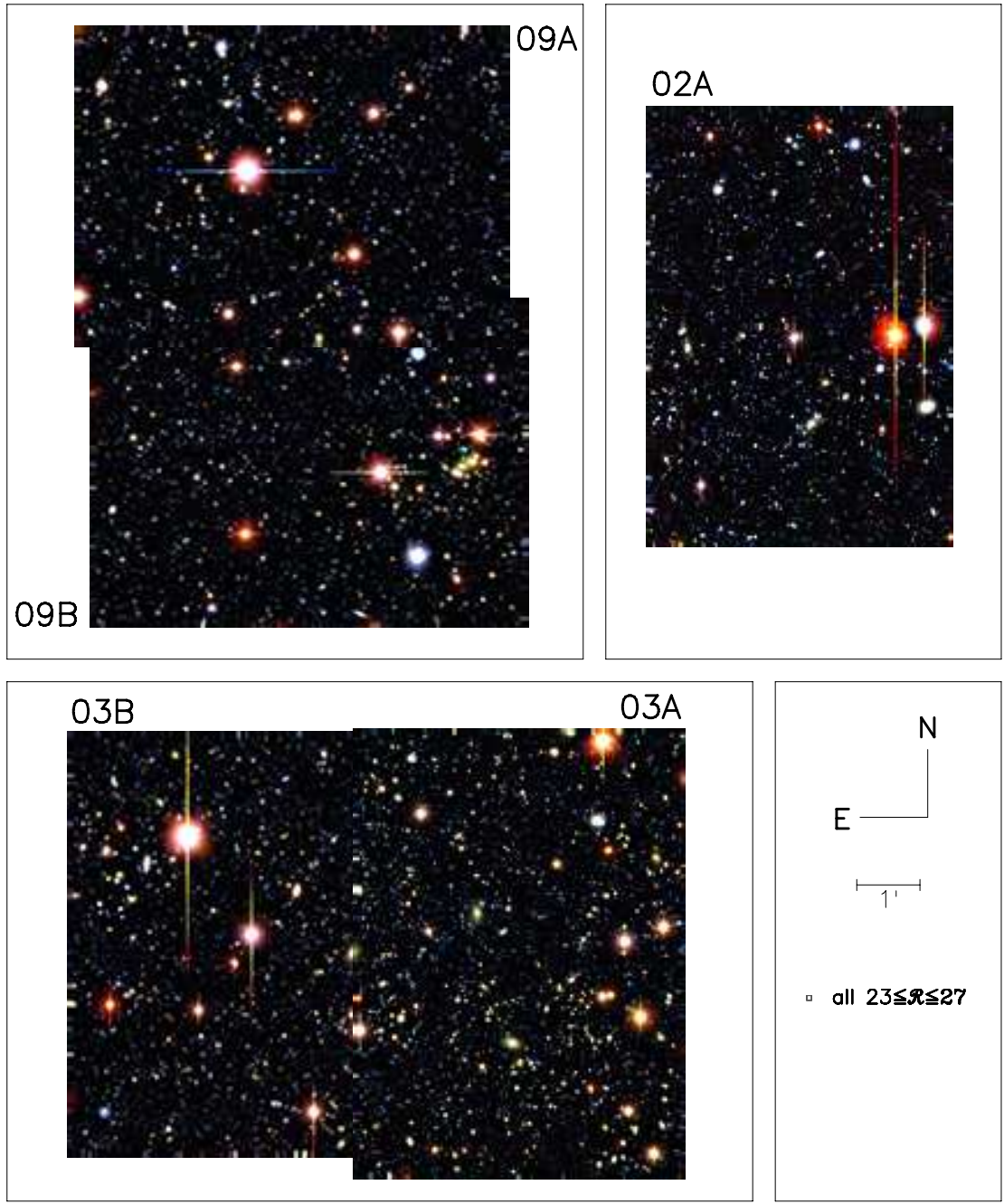}
\caption{\label{color-images.fig} 
A HIGH RESOLUTION VERSION OF FIGURE 1 CAN BE OBTAINED FROM THE
AUTHORS.  Composite \GRI\ color images of the KDF. \I-band is shown as
red, \R-band as green, and \G-band as blue.  Although we have made a
reasonable attempt at colormap consistency between the fields, small
systematic color differences between the fields in this figure may
remain.  Therefore, apparent differences between colors of objects
from field to field in this figure should not be overinterpreted.  }
\end{figure}

The \GRI\ color composite images shown in
Figure~\ref{color-images.fig} give a visual overview of the KDF, while
Table~\ref{field-details.tab} summarizes key information about our
survey fields, including field coordinates and sizes, foreground
extinction, exposure times per filter, image quality, and limiting
depth.  As shown in Fig.~\ref{color-images.fig}, the survey consists
of five LRIS fields, two pairs of which are abutting along their long
edges.  The survey area is thus split into two larger 'patches' of two
LRIS fields each and a third, smaller patch that consists of a single
LRIS field.  Our field-naming convention is based on the right
ascensions of the patches and we call the five fields 02A, 03A and 03B
(together comprising the 03 patch), and 09A and 09A (the 09 patch).
The three patches are widely separated on the sky, helping to ensure
that the effects of large-scale structures are averaged out.  The
purpose of abutting pairs of fields into the larger patches is to give
us improved ability to study galaxy clustering (Sawicki \& Thompson,
in preparation), where the larger {\it contiguous} area is important
as it gives us many more baselines in general and, particularly, many
long baselines few of which are available in a single LRIS field.  Our
09 patch partially overlaps one of the \UGRI\ fields (field Q0933+289)
from the survey of Steidel et al.\ (2003); this patch contains a
$z$=3.43 quasar (the brightest object in the 09A field --- see
Figure~\ref{color-images.fig}), but we do not expect its presence to
bias our science as there is no evidence of a galaxy overdensity at
the QSO redshift in the extensive Steidel et al.\ (2003) spectroscopy
of this field.

Our survey was carried out in dark-time over two three-night observing
runs: 2001 December 18--20 UT (hearfter Run~1) and 2002 December 2--4
UT (hereafter Run~2).  All data were taken using the LRIS imaging
spectrograph (Oke et al.\ 1995; McCarthy et al.\ 1998; Steidel et al.\
2003) mounted on the Keck I telescope.  LRIS provides two independent
channels, fed through a dichroic beamsplitter, which {\it
simultaneously} observe two separate wavelength ranges.  We used the
D560 dichroic, which allowed us to observe simultaneously in \U\ (or
\G) on the blue side and \R\ (or \I\, or $Z$) on the red side.  The
red channel uses a SITe/Tektronix 2048$\times$2048 pixel,
backside-illuminated CCD with 0.215\arcsec\ pixels and a quantum
efficiency that peaks at 70\% at 6000\AA\ and remains above 50\% to
8500\AA.  The detectors in the blue channel changed half-way through
the project: during Run 1 we had to use a fairly average SITe
2048$\times$2048 pixel CCD (QE$\sim$35\% at \U\ and $\sim$65\% at \G),
but during Run 2 we benefited from an excellent mosaic of two
UV-optimized EEV (Marconi) 2048$\times$4096 pixel CCDs with
0.135\arcsec\ pixels and very high UV and blue quantum efficiency
(QE$>$60\% at \U\ and $>$85\% at \G).  In all these detectors dark
current was negligible.

We lost the bulk of the first night of Run 1 to weather and instrument
problems, but the remaining two nights of Run 1, as well as all three
nights of Run 2 were both photometric and trouble-free.  
\R-band seeing ranged over 0.7--1.4\arcsec, although we rejected the 
frames with the poorest image quality when making our stacked images
(\S~\ref{datareductions}).  During part of one night of Run 1 the
seeing was particularly poor, affecting all the \G-band images of one
of the fields (field 03A). Consequently, all the Run~1 \G-band images
of this field was rejected and was redone in Run~2 while at the same
time some additional \I-band imaging of that field was simultaneously
obtained using the red channel of the instrument.

The data were acquired as a series of short, dithered exposures.  The
\U-band frames were always acquired simultaneously with the \R-band
frames, and the \G-band frames with the \I-band ones. Aditionally, a
small amount of $Z$-band data, taken through the long-pass RG850
filter, were taken in parallel with the \U-band observations; these
$Z$-band data will be presented elsewhere.  Individual exposure times
were 1200s in \U-band, 1200s in \G-band, 525s (or 585s) in
\R-band, and 325s (or 360s) in \I-band.  Exposure times in the red
channel (\R\ and
\I\ bands) --- where the night-time sky is quite bright even during
darktime --- were short to avoid straying into the non-linear regime
of the CCD response.

We wished to dither but did not want to waste time not taking data in
one channel while waiting for the other channel to finish.
Consequently, we set the red channel exposure times so that the end of
the last read-out of a set of red exposures coincided with the end of
the read out of a blue exposure:
\begin{equation}
N_{exp}^{red}\times(t_{exp}^{red}+t_{read}^{red})= (t_{exp}^{blue}+t_{read}^{blue}),
\end{equation}
where $t_{exp}$ are the exposure times, $t_{read}$ are the read-out
times, and $N_{exp}^{red}$ is the number of red channel exposures
taken during one blue-channel exposure ($N_{exp}^{red}$=2 for \R-band
and 3 for \I-band).  After each such exposure sequence consisting of
one blue-channel and $N^{red}_{exp}$ red-channel exposures, the
telescope was offset following a quasi-random pattern on a nine-point
square grid with 24\arcsec or 30\arcsec edges.  This dithering was
done to allow the subsequent exclusion of bad pixels from the stacked
frames and to permit the masking-out of sources in the construction of
flat-field images and fringe frames from the imaging data.

In total, our good \UGRI\ images of the five fields consumed 71.1
hours of {\it exposure time}. This number does not include overheads,
images that were excluded from the final stacks because of poor image
quality etc., nor the $Z$-band data.  Once processed and stacked,
these images comprise some of the deepest multiwavelength imaging
taken, particularly the U-band, over this wide an area.

\section{DATA REDUCTION}\label{datareductions}

\subsection{Pre-processing and frame stacking}

The data were processed in IRAF\footnote{IRAF is distributed by the
National Optical Astronomy Observatories, which are operated by the
Association of Universities for Research in Astronomy, Inc., under
cooperative agreement with the National Science Foundation.} using a
fairly standard algorithm for processing CCD data.  The primary
deviation is embodied in the iterative determination of the
fringe-correction image and the corrected domeflat.  We reduced each
dataset from a different CCD separately and, where possible, we also
independently processed data from separate nights.

In particular, the initial processing of the data consisted of first
subtracting the bias signal from each image using the overscan
regions.  Separate bias frames were then stacked and subtracted from
the science data to correct for any residual bias structure.  We then
constructed normalized domeflats in each filter, rescaling the images
to a constant mean value and stacking them with sigma-clipping to
remove cosmic rays.  We constructed a bad pixel mask from the bias and
domeflat images, marking as bad any hot or dead pixels and bad columns
or charge traps.

After this initial preparation, we flatfielded the science data with
the domeflats, then rescaled and stacked the images to reject real
sources on the sky.  We used these stacked images to construct a
fringe frame for the \G, \R, and \I\ data (the \U\ data did not need
fringe correction).  Any residual gradients were used to correct the
domeflats, and these few steps repeated until we converged on a good
fringe frame and corrected domeflat.  The final reduction of the
science data produced individual flat-fielded, fringe-corrected images
with zero mean sky values.

We then shifted the images (using integer pixle shifts) to correct for
dithering offsets and stacked all of the data for each field and
filter combination to produce preliminary deep images.  The deep
images were chopped, scaled, and convolved as necessary to subtract
from the individual images to aid in identifying and masking satellite
or meteor trails, asteroids, and cosmic rays.  Instrumental magnitudes
for a set of isolated, unsaturated sources in common to all frames
were determined and used to multiplicatively rescale the images to
constant photometry.  We measured the sky noise ($\sigma_i$) and
full-width at half-maximum ($FWHM_i$) of the seeing in each of these
images, $i$, and then produced the final deep image with a modified
variance weighting, which is corrected for the seeing:
\begin{equation}\label{weights.eq}
weight_i \propto 1 / (\sigma_i^2 FWHM_i).
\end{equation}
This scheme gives higher weight to the images with the best seeing
and/or the lowest noise, resulting in a gain in the depth of the
final, stacked images.

\subsection{Image registration, trimming, and photometric calibration}
\label{match-calibrate}

Next, the stacked \U, \G, and \I\ images were geometrically
transformed to match their corresponding \R-band images using the
positions of multiple bright point sources.  These aligned
\UGRI\ images of each field were then trimmed to exclude areas of low
S/N around the edges that were a byproduct of spatial dithering; this
trimming was done so that in each field the trimmed images in each of
the four bands covered only a common area, with all low-S/N edge areas
trimmed out.  Even after this trimming, sufficient overlap remained
between the 09A and 09B images (5.0 \sqam) and the 03A and 03B images
(1.8 \sqam) to allow us to later accurately tie together the two
fields of each patch.

The images were then photometrically calibrated.  We tied our
photometric system to photometrically calibrated deep \UGRI\ images
that were kindly provided to us for that purpose by Chuck Steidel.
Although the filter sets used in Steidel's images and in ours were
nearly identical, we nevertheless computed the invariably small color
terms in our photometric transformations since we wanted to be able to
replicate without bias the Steidel et al.\ (1998, 2003, 2004)
color-color selection of high-$z$ galaxies.  The use of these color
terms, derived from the comparison of images obtained using our
apparatus with those taken by Steidel et al.\ in their surveys,
ensures that any wavelength-dependent differences due to detector QE,
mirror reflectivity, etc., are calibrated out, and that we are working
on the same photometric system as Steidel et al. are.  As the final
step in our photometric calibration, we have corrected for the very
small effect of foreground Galactic dust extinction as determined from
the Schlegel, Finkbeiner, \& Davis (1998) dust maps.  The \Ebv\ values
of these dust corrections are given in Table~\ref{field-details.tab}.

In addition to the natural-seeing images, we also produced
seeing-matched \UGRI\ images for each field.  These seeing-matched
images were made for use in measuring object colors
(\S~\ref{objectfinding}) and were made by smoothing the three images
of the field that have better seeing to match the seeing in the
fourth, poorest-quality image.  The smoothing was done using a
Gaussian kernel whose width was determined for each image based on the
sizes of multiple bright but unsaturated point sources.  The FWHM
sizes of the seeing in the resultant smoothed images are given as
``common smoothed seeing'' in Table~\ref{field-details.tab}).  They
are typically FWHM \s 1.0--1.1\arcsec, except for the 03A field for
which the final image quality was significantly worse at 1.4\arcsec.
Our \s~1\arcsec\ seeing is very comparable to the seeing in the
shallower \UGRI\ surveys of Steidel et al.\ (1999, 2003, 2004).

\subsection{Object detection and photometry}\label{objectfinding}

Overall, our photometric approach is very similar to that used by
Steidel and collaborators, including the use of the same \UGRI\ filter
set, \R-band detection of objects, and color measurement through
circular apertures.  The specifics are discussed below and any
differences in approach are noted.

Ideally, we would have wished to {\it detect} galaxies at a constant
{\it rest-frame} wavelength irrespective of redshift. Given our data
and redshift ranges of interest this could have been done at
$\lambda$$\approx$1700\AA\ which corresponds to observed \I-band at
\zs4, \R-band at \zs3, and \G-band at \zs2.  However, because
our \R-band images are extremely deep, we chose to do object detecion
in the \R-band images only.  This approach is completely analogous to
the procedure used by the Steidel team and has the virtue of
simplicity in that it results in only one source catalog per field.
Given the depth of our \R-band images and the relatively mild
\RI\ and \GR\ colors of \zs4 and \zs2 galaxies, respectively (see 
Figs.~\ref{GRIcolorcolor.fig} and \ref{UGRcolorcolor.fig}, and also
Steidel et al.\ 1998, 2003, 2004), our \R-band object detection should
not bias our samples in any significant way.

We used the SExtractor package (Bertin \& Arnouts, 1996) for object
detection and photometry.  To detect objects, we ran SExtractor on the
{\it unsmoothed} \R-band images.  Total \R-band magnitudes are
SExtractor's MAG\_AUTO apertures which are Kron-like (Kron 1980)
elliptical apertures in which fluxes are corrected for any
contaminating close companions through masking and image
symmetrization.  Colors were measured in Sextractor's dual image mode,
using the unsmoothed images for object detection and the smoothed,
seeing-matched images for color photometry.  In keeping with the
approach of the Steidel team, we used 2.0\arcsec-diameter circular
apertures (Steidel et al.\ 2003) for color measurement.  Finally, the
instrumental magnitudes were transformed onto the AB system (Oke 1974)
using the zeropoints and color terms determined in
\S~\ref{match-calibrate}.  The total magnitudes in the
\U, \G, and \I\ bands were then straightforwardly computed from the 
\R-band total  magnitudes and aperture colors via
\begin{equation}
m_{tot}={\cal R}_{tot}-({\cal R}_{ap}-m_{ap}),
\end{equation}
where \R\ is the \R-band magnitude, $m$ represents the magnitude in
one of the other bands (\U, \G, or \I) and the subscripts $ap$ and
$tot$ denote aperture and total magnitudes, respectively.

Our treatment of ``drop-out'' objects also follows closely the
approach of Steidel et al.\ (2003).  If an object's reported flux in
the \U\ (or \G) band is greater than the 1$\sigma$ fluctuation in the
sky background over the size of our color aperture then the object is
considered detected in that band and is assigned an aperture magnitude
that corresponds to that measured flux.  If, on the other hand, the
reported flux is below the 1$\sigma$ threshold, the object is
considered a drop-out and is assigned an upper flux limit that is
defined as the magnitude that corresponds to the 1$\sigma$ sky
fluctuation.

The main difference between our procedure and that of Steidel and his
team is that whereas they use an in-house modified version of FOCAS
for object detection and photometry, we use SExtractor.  The use of
these two different programs has two potential consequences. 
\begin{enumerate} 

\item {The first of these potential consequences is that the number of 
objects detected by SExtractor and FOCAS may be different as the two
programs use different object-detection algorithms.  If these were the
case than we might expect a different number density of galaxies in
the two surveys.  However, this is {\it not} the case here:
Table~\ref{field-details.tab} lists the number densities of objects
with \R=22.5--25.0 in our fields. These number densities are entirely
consistent with the distribution of identically defined object
densities in the 17 fields of Steidel et al.\ (2003) and the average
of our five fields, $<$N$>$=27.7$\pm$2.9, is fully consistent with
their 17-field average of $<$N$>$=25.9$\pm$1.9.  We therefore conclude
that any differences in object-finding between FOCAS and SExtractor
are likely not significant for our purposes and result in variations
that are certainly smaller than field-to-field scatter due to cosmic
variance.}

\item {The second potential consequence of using SExtractor instead 
of FOCAS is that total fluxes measured by SExtractor may differ from
those measured by FOCAS. Such difference may arise because the two
programs use different ways of sky estimation and different
definitions of total aperture (SExtractor uses Kron-like apertures
corrected for near neighbor contamination, whereas FOCAS uses simple
padded isophotal magnitudes).  However, given the small angular sizes
of distant galaxies, the differences between FOCAS and SExtractor
should be small --- on order a few percent.  Moreover, any strong
differences would also also reflect in a discrepancy in the number
density of 22.5$\leq$\R$\leq$25 objects, and --- as we discussed above
--- no such discrepancy is seen.  Finally, in the most sensitive
respect --- that of determining object {\it colors} --- we follow a
recipe identical to that of Steidel and his team, measuring colors
through fixed, 2\arcsec-diameter apertures.  }

\end{enumerate}

Overall, our object selection procedure and photometry are thus very
similar to those used by Steidel et al.\ and we expect there to be no
significant systematic differences between the two approaches.

\begin{figure}
\plotone{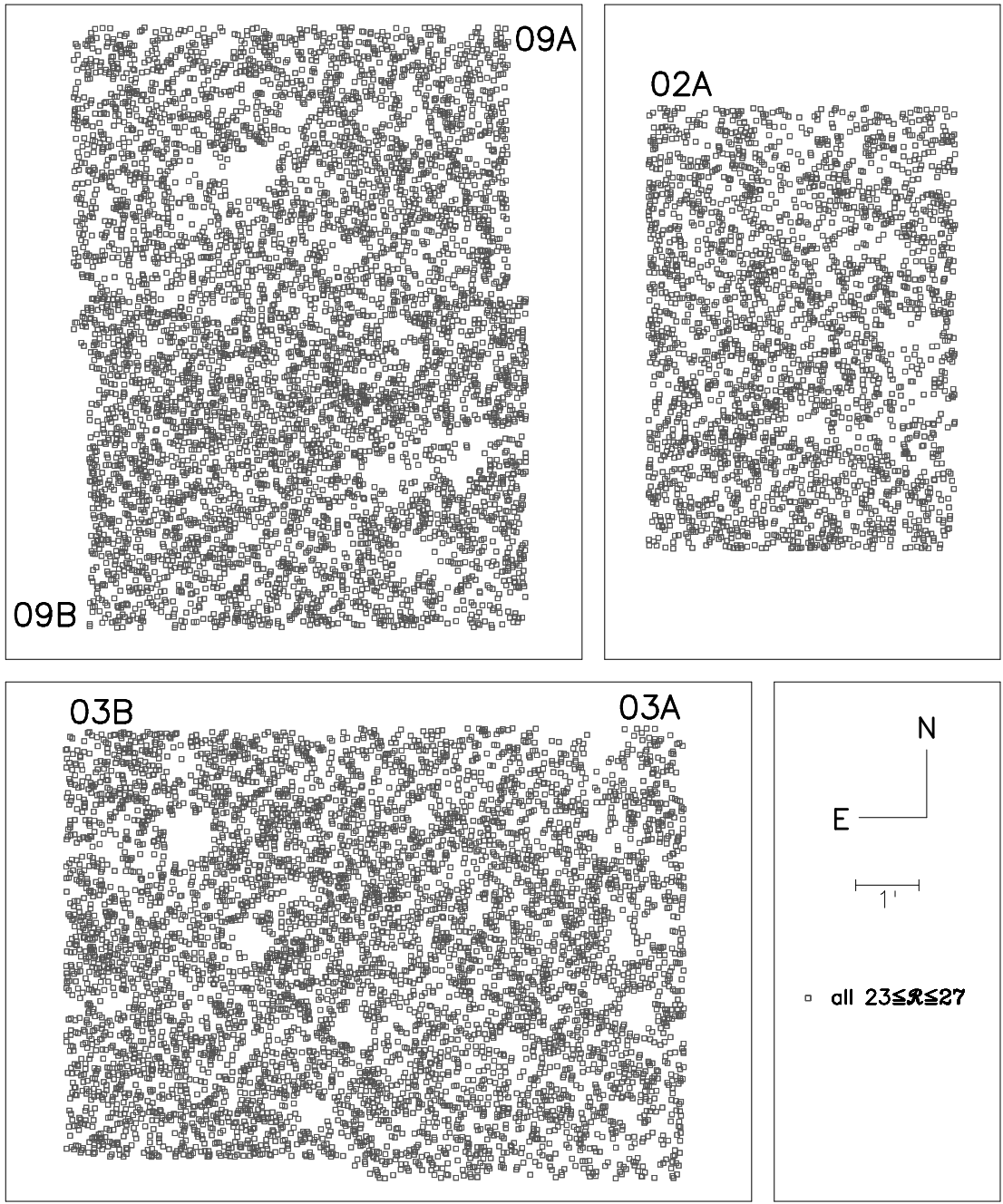}
\caption{\label{xypos.all.fig}
Positions of all objects with 23$<$\R$<$27. Note ``holes'' in which
object-finding was affected by light from bright foreground objects.}
\end{figure}

The spatial distribution of 23$\leq$\R$\leq$27 objects in our survey
is shown in Fig.~\ref{xypos.all.fig}.  Our catalog is clearly missing
objects in the vicinity of very bright stars (c.f.\
Fig.~\ref{color-images.fig}), where SExtractor has trouble finding
faint sources in the bright glow from the stars.  However, these areas
of low sensitivity are small and we will account for as needed using
simulations (\S~\ref{completeness}; also Sawicki \& Thompson 2004).
In all, our catalog contains 14579 objects with 23$\leq$\R$\leq$27 in
the 169~\sqam\ of our survey.

\subsection{Depth, Completeness, and Surface Brightness Selection Effects}
\label{completeness}

We assess the depth of the Keck Deep Fields in two ways: by measuring
the sky noise of our images and by conducting Monte Carlo simulations
that seek to detect artificial objects implanted into the images.

Table~\ref{field-details.tab} lists the sky surface brightness limits
measured from pixel-to-pixel RMS fluctuation in several
representative, object-free areas of each image.  In the redder bands
the image depth correlates closely with total exposure time, while in
the bluer bands it has a pronounced dependence on run-to-run changes
in detector sensitivity and on the sky brightness changes in the
individual exposures that were coadded into the stacked images.  In
the four fields that received close to the full intended exposure time
(fields 03A, 03B, 09A, and 09B) characteristic 1$\sigma$ sky surface
brightness limits are $\mu_{lim}$ $\sim$ 30.7, 30.8, 30.0, and 29.4
mag/\sqam\ in \UGRI. They are $\sim$0.5 mag shallower in the 02A
field.  In contrast, the typical sky surface brightness limits of the
imaging used by the Steidel team are 28.7, 29.0, 28.4, and 28.0 in
\UGRI\ (Steidel et al.\ 2003, 1999).  Thus, over the bulk of our
survey, we reach $\sim$1.5--2 mag deeper in sky noise than do Steidel
et al.  Even in our relatively shallow 02A field we reach sky surface
brightness limits that are 0.7--1.8 mag deeper than those in the
Steidel et al.\ surveys.

Sky surface brightness limits are of course only part of what affects
the sensitivity of faint object imaging.  A second key ingredient is
image quality, which is primarily influenced by seeing and is
parametrized by the amplitude of the stellar FWHM.  With the exception
of the 03A field, typical seeing of our stacked images is
$\sim$1\arcsec, which is very comparable to the seeing in the data
used by the Steidel team (Steidel et al.\ 2003).

We use Monte Carlo simulations to assess the combined impact of both
seeing {\it and} sky surface brightness on our data.  By their nature
these simulations also take into account other effects that impact the
detection rate, such as the poorer detection sensitivity around bright
stars and the confusion by close companions.  At this stage, we are
only interested in simply determining the {\it detection} efficiency
of faint objects in the \R-band KDF images and do not attempt here to
ascertain the incompleteness of our LBG sample due to scatter out of
color-color selection regions by photometric errors in object colors.
Such more comprehensive simulations are part of our study of the LBG
luminosity functions (Sawicki \& Thompson, in prep.).  

To gauge our detection efficiency, we made simulations that implanted
artificial objects with a range of fluxes and sizes at random position
into our images and which then attempted to recover them using the
same procedures as those we used in \S~\ref{objectfinding} to
construct our source catalogs.  To assess the sensitivity of our
catalog to object surface brightness, we simulated a range of
Gaussian-shaped sources with FWHM=0.5--2\arcsec.  However, we note
that HST imaging of LBGs shows them to be very compact, with
half-light radii $r_{1/2}$$\sim$0.1--0.3\arcsec\ over a range of
epochs: \zs5, \zs3, and \zs2 (Bremmer et al.\ 2004; Giavalisco,
Steidel, \& Macchetto 1996; Erb et al.\ 2003). Consequently, given
that our catalog is based on ground-based \R-band images with
FWHM$\geq$0.8\arcsec, our target high-$z$ galaxies are essentially
unresolved point sources with FWHM that corresponds to the seeing.

\begin{figure}
\plotone{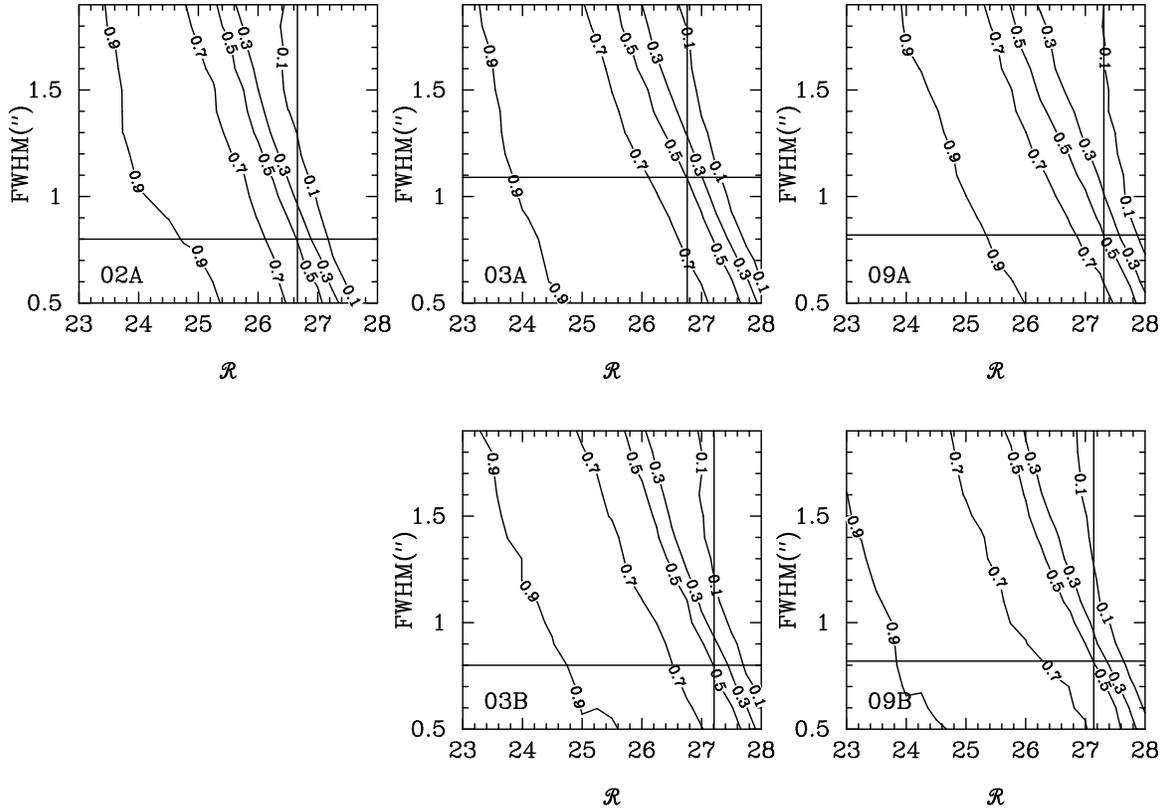}
\caption{\label{detection-efficiency.fig}
Completeness as a function of \R\ magnitude and object size. Contours
show completeness deremined from simulations (see text).  Object
compactness is parametrized as Gaussian FWHM --- as is appropriate for
intrinsically compact objects that are unresolved in our ground-based
images.  The point-source FWHM values are shown as horizontal lines.
High-$z$ star-forming galaxies are compact and virtually unresolved
under the $\geq$0.8\arcsec\ seeing in these images and therefore their
sizes are well approximated by the horizontal lines.  Our limiting
magnitudes, \Rlim, are defined as the magnitudes that correspond to
50\% detection efficiency for unresolved objects.  They are \Rlim\s27
and are shown with vertical lines.}
\end{figure}

The contours in Fig.~\ref{detection-efficiency.fig} show the results
of our detection-rate simulations.  The fraction of objects recovered
as a function of object FWHM and \R-band magnitude is shown for each
of the five fields. As expected, the detection efficiency drops with
increasing total magnitude and FWHM.  The horizontal line in each
panel shows the stellar FHWM as measured from several bright but
unsaturated stars in the images and correspond to the ``seeing (\R)''
values in Table~\ref{field-details.tab}.  As noted above, because of
the very small sizes ($r_{1/2}$$\sim$0.2\arcsec) of distant
star-forming galaxies, the horizontal line also reasonably represents
the expected FHWM of our targets.  As a measure of the detection depth
of our images we adopt the magnitude at which 50\% of unresolved
objects are detected (hereafter, \Rlim).  These limiting detection
magnitudes range over \Rlim=26.7--27.3 and are listed in
Table~\ref{field-details.tab} and shown as vertical lines in
Fig.~\ref{detection-efficiency.fig}.

Overall, three of our five fields (03B, 09A, and 09B) reach object
detection limits \Rlim\s27.2.  The other two fields, 02A and 03A,
reach \Rlim\s26.7.  The \s0.5 mag difference in depth arises because
the two shallower fields either had a shorter exposure time (02A) or
poorer seeing (03A) than the three deeper fields.

In summary, the detection limits, \Rlim, tell us to what depth we can
reasonably push our source list before encountering significant
detection incompleteness.  We find that \Rlim\s27.0 in the KDF, which
is \s1.5 magnitudes deeper than the work of the Steidel team.  The sky
surface brightness limits that were discussed earlier tell us how deep
we can push our matched-aperture color measurements without incurring
photometric errors that are larger than those in the surveys of
Steidel et al.  Here, we also found that we can do so to \s1.5
magnitudes deeper than is the case in the data of the Steidel team.
Thus, overall, our
\UGRI\ survey can select galaxies in a manner that's identical to that
used for the spectroscopically-calibrated selection of high-$z$
star-forming galaxies by the Steidel group, but can do so with
confidence for objects that are up to \s1.5 magnitudes fainter, namely
to \R\s27.0. It is to the selection of high-$z$ star-forming galaxies
that we now turn.

\section{PHOTOMETRIC SELECTION OF HIGH-$z$ GALAXIES}\label{LBGselection}

\subsection{Color-color selection criteria}\label{selection-criteria}

Steidel et al.\ (1999; 2003; 2004) have developed and extensively
tested color-color selection criteria that efficiently and robustly
select galaxies at \zs4, \zs3, \zs2.2, and \zs1.7.  These selection
criteria have evolved somewhat over time (c.f., e.g., Steidel et al.\
1996, Erb et al.\ 2003) and in our work we use the most recent
published selection criteria, which are as follows.

To select \zs4 objects we use (Steidel et al.\ 1999)
\begin{eqnarray}\label{z4sel.eq}
G-{\cal R} & \geq & 2.0, \nonumber \\
G-{\cal R} & \geq & 2({\cal R}-I)+1.5, \\
{\cal R} - I & \leq & 0.6, \nonumber 
\end{eqnarray}
for \zs3 objects we use (Steidel et al.\ 2003)
\begin{eqnarray}\label{z3sel.eq}
G-{\cal R} & \leq & 1.2, \nonumber \\
U_n - G & \geq & G - {\cal R} + 1.0, \\
G-{\cal R} & \geq & -0.1, \nonumber
\end{eqnarray}
for \zs2.2 objects we use (see Steidel et al.\ 2004)
\begin{eqnarray}\label{z22sel.eq}
G-{\cal R} & \geq & -0.2,  \nonumber\\ 
U_n-G & \geq & G-{\cal R}+0.2,  \\
G-{\cal R} & \leq & 0.2(U_n-G)+0.4,  \nonumber\\ 
U_n-G & \leq & G-{\cal R}+1.0,\nonumber
\end{eqnarray}
and for \zs1.7 objects we use (Steidel et al.\ 2004)
\begin{eqnarray}\label{z17sel.eq}
G-{\cal R} & \geq & -0.2,  \nonumber\\ 
U_n-G & \geq & G-{\cal R}-0.1,  \\
G-{\cal R} & \leq & 0.2(U_n-G)+0.4,  \nonumber\\ 
U_n-G & \leq & G-{\cal R}+0.2.\nonumber
\end{eqnarray}
Additionally, we impose a faint magnitude limit of \R$\leq$27.0
motivated by the depth of our images. To guard against bright
foreground interlopers we also impose a bright-end cut of
\R$\geq$23.0.

\begin{figure}
\plotone{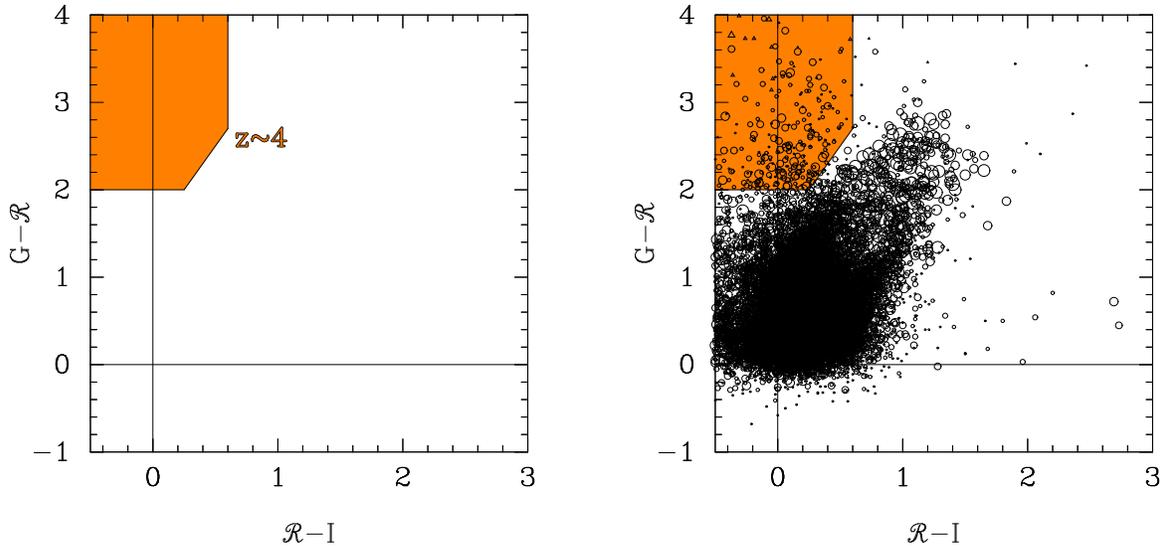}
\caption{\label{GRIcolorcolor.fig} 
\GRI\ color-color plots.  The redshift selection region defined by 
Equation \ref{z4sel.eq} is highlighted in orange; the filters and
color-color selection used are identical to those used by Steidel et
al.\ (1999) to select \zs4 galaxies.  The colors of objects with
23$\leq$\R$\leq$27 are shown in the right-hand panel.  Circles denote
objects that were significantly detected in all bands, and
upward-pointing triangles show objects with \G-band upper flux limits
only. Symbol size corresponds to \R\ magnitude.  Note the close
similarity of the color-color distribution in this figure to that in,
e.g., Fig.~2 of Steidel et al.\ (1999). }
\end{figure}

\begin{figure}
\plotone{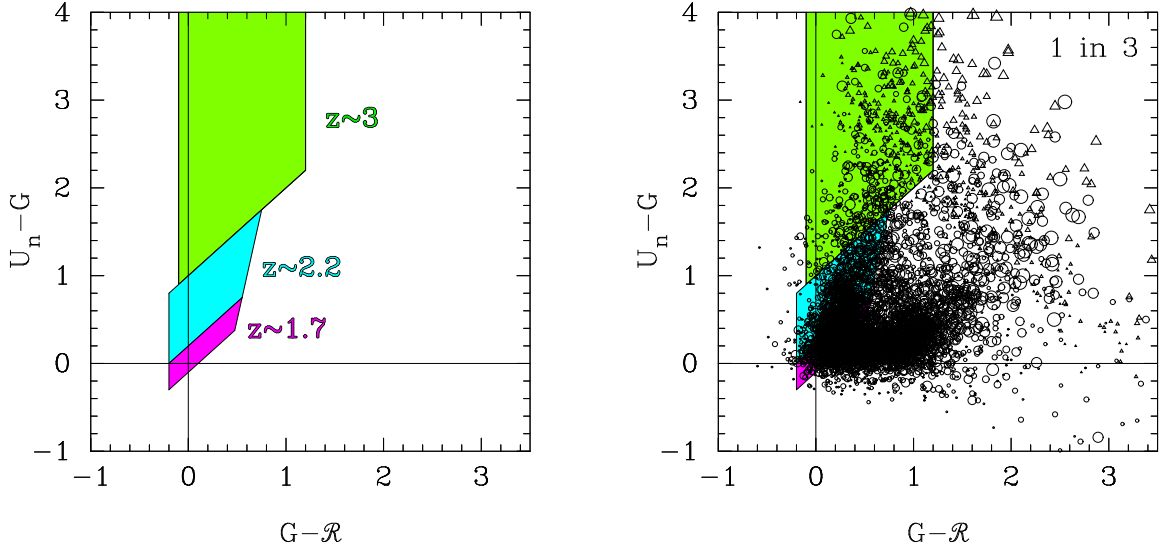}
\caption{\label{UGRcolorcolor.fig} 
\UGR\ color-color plots.  The left panel shows redshift selection 
regions defined by equations \ref{z3sel.eq} (green),
\ref{z22sel.eq} (blue), and \ref{z17sel.eq} (magenta).  These regions 
are {\it identical} to the well-known and well-characterized selection
criteria of Steidel et al.\ (2003, 2004).  The right panel shows the
colors of our objects with 23$\leq$\R$\leq$27.  For clarity, only one
object out of three is plotted.  Circles denote objects that were
significantly detected in all bands, and upward-pointing triangles
show ubjects with \U-band upper limits only.  Symbol size corresponds
to \R\ magnitude.  Note the close similarity of the color-color
distribution of objects in this figure to that in, e.g., Fig.~1 of
Steidel et al.\ (2004). }
\end{figure}

The color-color selection criteria of
Equations~\ref{z4sel.eq}--\ref{z17sel.eq} are illustrated in the
left-hand panels of Figures~\ref{GRIcolorcolor.fig} and
\ref{UGRcolorcolor.fig}. The left panel of Figure~\ref{GRIcolorcolor.fig} 
shows the region in \GR\ vs. \RI\ color space used to select galaxies
at \zs4 (Eq.~\ref{z4sel.eq}).  The left panel of
Figure~\ref{UGRcolorcolor.fig} shows in green, blue, and magenta the
regions of \UG\ vs.\ \GR\ color-color space defined by
Equations~\ref{z3sel.eq}, \ref{z22sel.eq}, and \ref{z17sel.eq} (\zs3,
2.2, and 1.7), respectively.

The criteria of Eq.~\ref{z3sel.eq} correspond exactly to the union of
LBG types C, D, M, and MD of Steidel et al.\ (2003); those of
Equations~\ref{z22sel.eq} and \ref{z17sel.eq}, respectively, to what
Steidel et al.\ (2004) call types BX and BM.  In our work we do not
use this nomenclature of Steidel et al., but --- motivated by the
observed redshift distributions (see below) of objects selected by
Equations~\ref{z4sel.eq}--\ref{z17sel.eq} --- refer to them as the
``\zs4'', ``\zs3'', ``\zs2.2'', and ``\zs1.7'' criteria.

Extensive spectroscopy of hundreds of objects (Steidel et al.\ 1999,
2003, 2004) has shown that the redshift distributions of objects
selected by the criteria of Equations~\ref{z4sel.eq}--\ref{z17sel.eq}
are --- at least for their shallower samples --- roughly
Gaussian-shaped with $<$$z$$>$=4.13$\pm$0.26, $<$$z$$>$=2.96$\pm$0.26,
$<$$z$$>$=2.20$\pm$0.32, and $<$$z$$>$=1.70$\pm$0.34.  Spectroscopy
has also shown that there is only small contamination by Galactic
stars, low-$z$ galaxies, or high-$z$ AGN.  At intermediate magnitudes,
\R\s24--25.5, the contamination by foreground interlopers --- defined
as objects with $z$$<$1 --- is less than \s5\% in all three
\UGR-selected redshift bins (\zs1.7, 2.2, and 3; Steidel et al.\ 2003,
2004) and is likely to be even smaller in our samples because the
ratio of galaxies to Galactic stars increases towards fainter
magnitudes.  The AGN fraction is put at $\sim$3\% by Steidel et al.\
(2003, 2004).  In the
\GRI-selected \zs4 sample, the foreground contamination is somewhat
higher, \s20\%, although the statistics are fairly poor due to the
small numbers of
\GRI-selected objects with spectroscopy (Steidel et al.\ 1999).

In summary, the color-color selection criteria of
Equations~\ref{z4sel.eq}--\ref{z17sel.eq} select distinct populations
with fairly narrow spreads in redshift of $\delta$$z$$\sim$$\pm$0.3
and with very little contamination by interlopers.  We now turn to
apply these well-understood selection criteria to our data.

\subsection{Our high-$z$ galaxy sample}

The right-hand panel of Fig.~\ref{GRIcolorcolor.fig} shows the \GR\
vs.\ \RI\ colors of the 23$\leq$\R$\leq$27 objects in the KDF.  The
right-hand panel of Figure~\ref{UGRcolorcolor.fig} shows the \UG\ vs.\
\GR\ colors of 23$\leq$\R$\leq$27 objects, although --- for clarity
--- only 1 in 3 objects are plotted.  The color distributions of
objects in these two figures are very similar to the corresponding
brighter samples of the Steidel et al.\ surveys (see Fig.~2 of Steidel
et al.\ 1999 and Fig.~1 of Steidel et al.\ 2003).  This close
similarity is not surprising given the similarity of the KDF image
depths at \R\s27 to theirs at \R\s25.5.  However, it does gives us
strong reassurance that we are selecting identical populations, with
similar photometric scatter, though at substantially fainter
luminosities.

To \R=27.0, the color-color selection gives us 427 \zs4, 1481 \zs3,
2417 \zs2.2, and 2043 \zs 1.7 star-forming galaxies in the 169 \sqam\
of the KDF.  This gives surface densities of, 2.5, 8.8, 14.3, and
12.1~arcmin$^{-2}$ at \zs4, \zs3, \zs2.2, and \zs1.7, respectively.
These densities are significantly higher than the surface densities of
identically-selected but brighter objects in the \R$\leq$25.5 samples
of Steidel et al.\ (1999, 2003, 2004), which is not surprising in
light of the fact that the KDF probe considerably deeper into the
faint end of the luminosity function at these redshifts.  We discuss
in detail the shape and evolution of the high-$z$ galaxy luminosity
function in a separate paper (Sawicki \& Thompson, in prep.).

\begin{figure}
\plotone{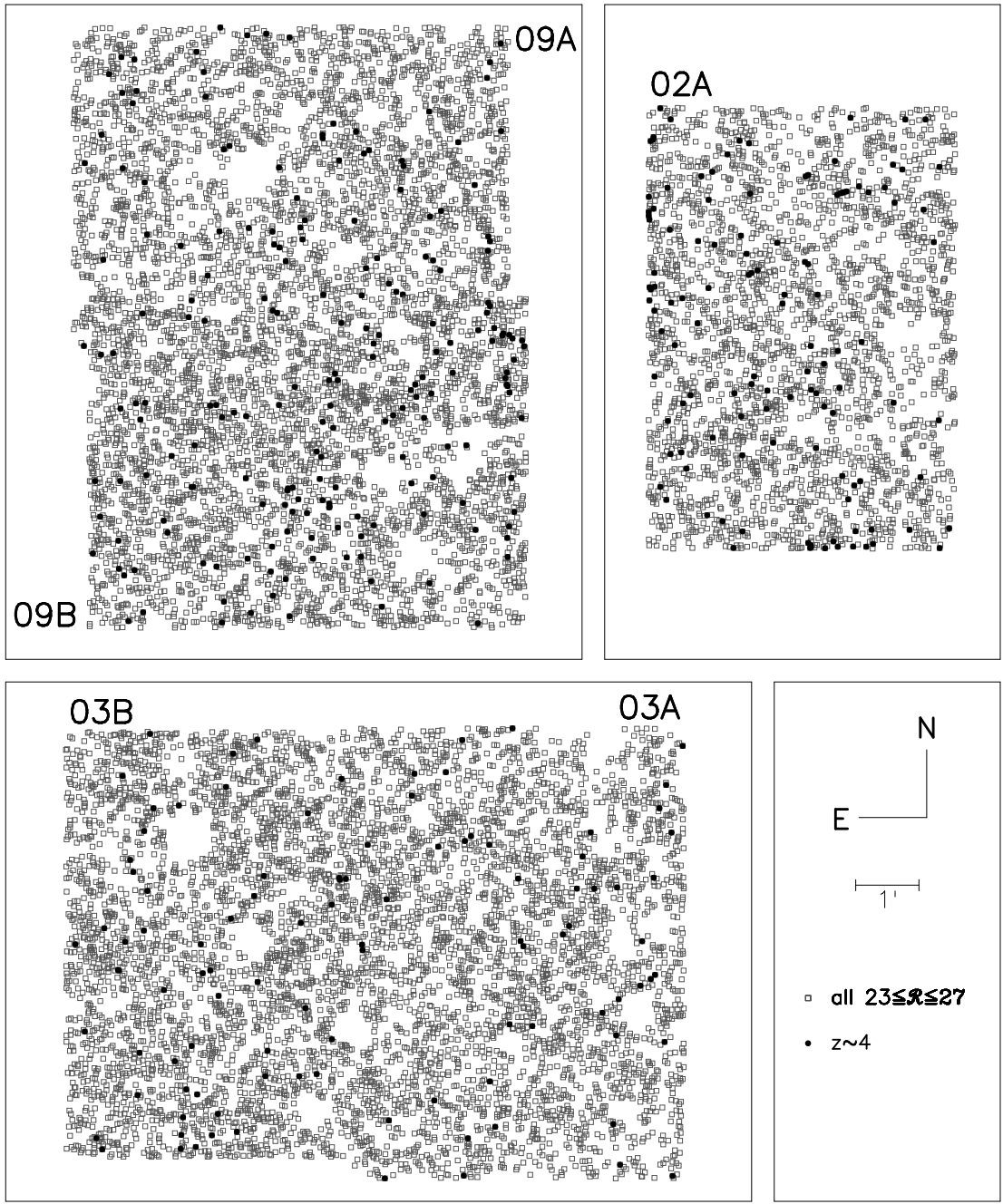}
\caption{\label{xypos.z4.fig}
Positions of \zs4 color-selected star-forming galaxies overplotted on
top of the positions of all 23$\leq$\R$\leq$27 objects in the
survey. }
\end{figure}

\begin{figure}
\plotone{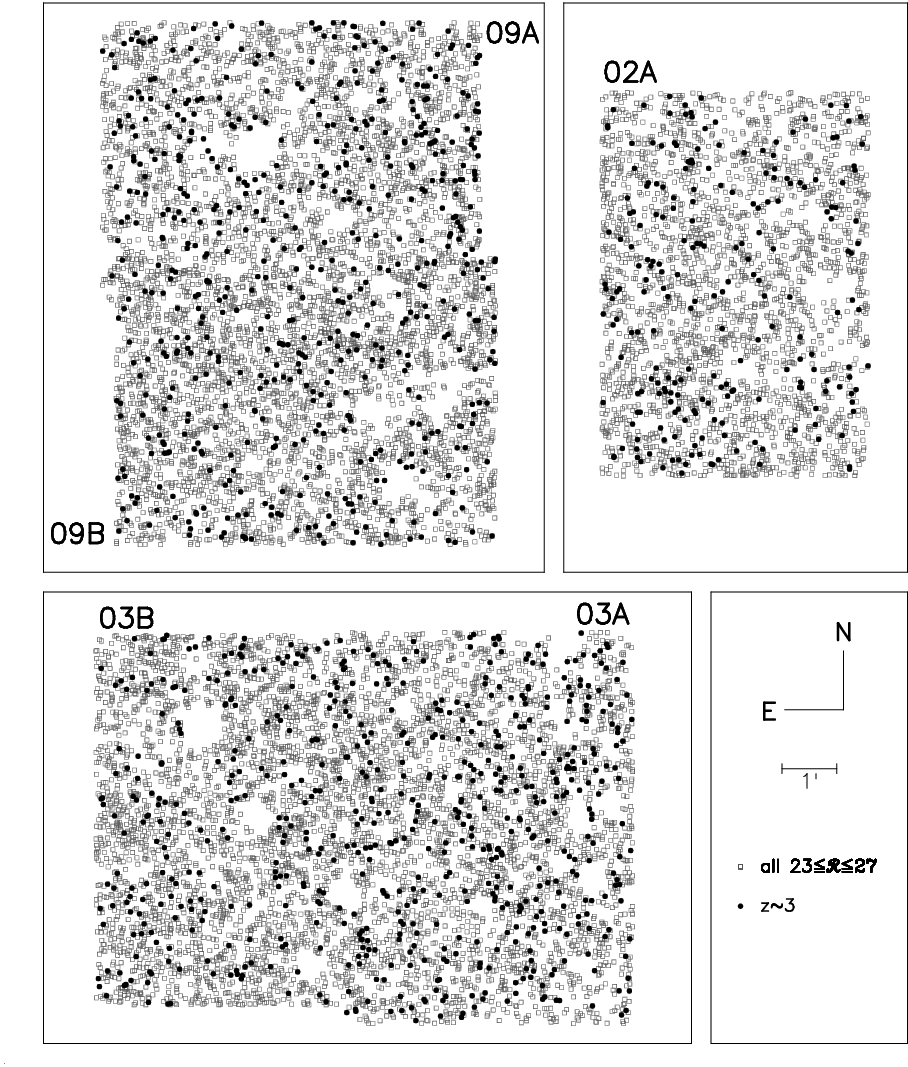}
\caption{\label{xypos.z3.fig}
Positions of \zs3 color-selected star-forming galaxies. }
\end{figure}

\begin{figure}
\plotone{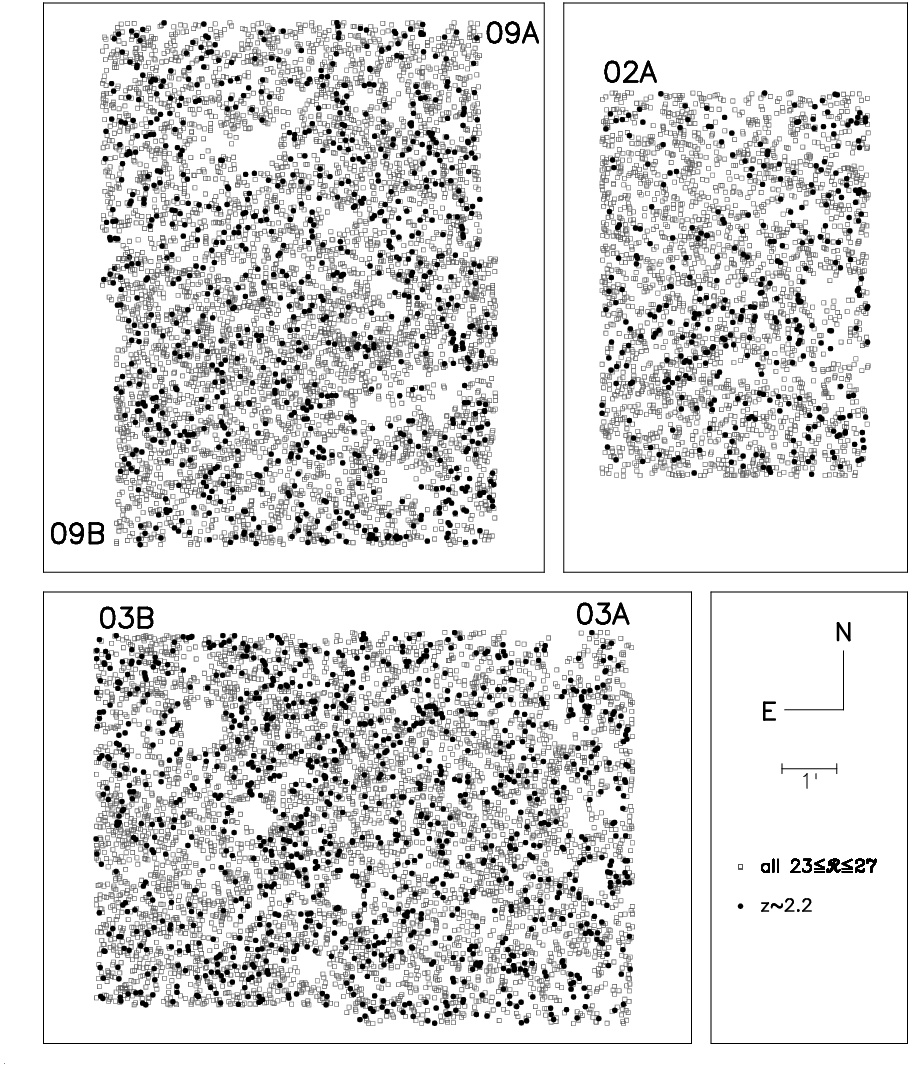}
\caption{\label{xypos.z22.fig}
Positions of \zs2.2 color-selected star-forming galaxies. }
\end{figure}

\begin{figure}
\plotone{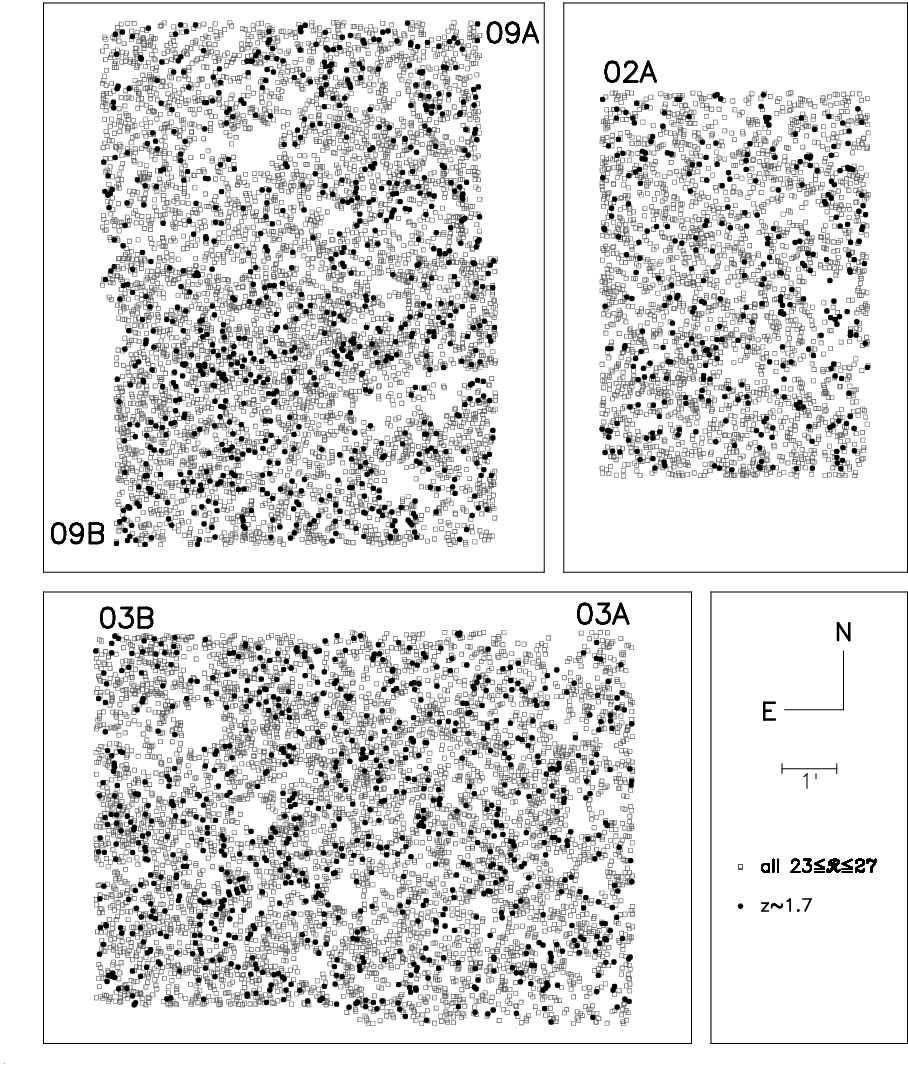}
\caption{\label{xypos.z17.fig}
Positions of \zs1.7 color-selected star-forming galaxies. }
\end{figure}

Figures~\ref{xypos.z4.fig}--\ref{xypos.z17.fig} show the spatial
positions of the color-color selected objects in our survey
overplotted on the positions of all
\R-selected objects.  As with Fig.~\ref{xypos.all.fig}, there are
detection ``voids'' in the vicinity of bright stars (c.f.\
Fig.~\ref{color-images.fig}).  Additionally, however, the high-$z$
galaxies shown in these four redshift slices do show significant real
clustering: numerous voids and overdensities can be seen in all four
redshift slices, and there are also hints of filaments, best seen in
the \zs4 sample in Fig.~\ref{xypos.z4.fig}.  High-$z$ galaxies are, of
course, well known to cluster, (e.g., Adelberger et al.\ 1998;
Giavalisco et al.\ 1998; Ouchi et al.\ 2001), and their clustering
can, for example, be used to constrain the properties of the dark
matter halos that they inhabit.  Because of its depth, area, and large
redshift span, our KDF sample is uniquely well suited to the study of
clustering evolution and its dependence on luminosity.  We will study
these issues in detail in Sawicki \& Thompson (in prep.).

\subsection{Contamination and completeness of the high-$z$ samples}
\label{contamination-completeness}

As is the case in any color-color selection of high-$z$ galaxies, our
high-$z$ samples may suffer both from selection bias and from
contamination by foreground interlopers.  Four effects can be at play:
some high-$z$ galaxies may be missed because they have {\it intrinsic}
colors that are outside the color-color selection regions defined by
Equations~\ref{z4sel.eq}--\ref{z17sel.eq}; low-$z$ interlopers may be
included because they have intrinsic colors inside the color-color
selection regions; high-$z$ galaxies with intrinsic colors that are
inside the selection regions may scatter out of them due to
photometric errors; and, finally, foreground interlopers may scatter
into the selection regions due to photometric errors.  These four
effects will affect our completeness by making us miss some fraction
of high-$z$ galaxies from our sample, yet will also contaminate our
sample with foreground interlopers.  We discuss the importance of
these four effects in turn, making particular use of the {\it
spectroscopically constrained} contamination fractions of the Steidel
et al.\ surveys.

\begin{enumerate}

\item {\it Low-$z$ objects with intrinsic colors that place 
them in the high-$z$ samples.}  While the color-color selection
criteria of Equations~\ref{z4sel.eq}--\ref{z17sel.eq} are very
effective at selecting high-$z$ galaxies from the much more numerous
foreground objects, they are not immune against low-$z$ objects whose
{\it intrinsic} colors place them in the high-$z$ color-color
selection boxes.  The colors of certain types of Galactic stars, for
example, put them into the regions of
Equations~\ref{z3sel.eq}--\ref{z17sel.eq}, and the colors of $z$\s1
red galaxies come dangerously close to the $z$$\sim$4 selection box
defined by Equation~\ref{z4sel.eq} (see Steidel et al.\ 1999).
Ordinarily, we would have no robust way of determining interloper
fractions without recourse to {\it very} expensive spectroscopy.
However, because we use the very same \UGRI\ filters and color-color
selection criteria as the Steidel group, we can use their
spectroscopically-determined contamination fractions to constrain the
fraction of such interlopers in our samples.  At \R\s25, the
interloper fractions are $\lesssim$5\% for \zs1.7, 2.2, and 3, and
\s20\% at \zs4 (Steidel et al. 1999, 2003, 2004; see also
\S~\ref{selection-criteria}).  Most of these interlopers are Galactic 
stars and intermediate-redshift ($z$\s1) red galaxies.  However, at
the magnitudes of our survey, the interloper fractions should be lower
than in the surveys of Steidel et al.\ because the ratio of galaxies
to Galactic stars decreases at fainter magnitudes as one ``punches''
out of the Galaxy, and --- similarly --- the fraction of
intermediate-$z$ red galaxies decreases as one probes past the peak of
their luminosity function.  Thus, we can expect that the interloper
fractions measured by Steidel et al.\ at \R\s25 are {\it higher} than
the interloper fractions at the fainter magnitudes of our survey.  We
therefore conservatively conclude that the interloper fractions in our
\zs1.7, 2.2, and 3 samples are $\lesssim$5\%, and are  $\lesssim$20\% 
at \zs4.

\item {\it Low-$z$ objects scattered into the high-$z$ samples by photometric errors.} 
In addition to low-$z$ interlopers whose {\it intrinsic} colors lie in
the high-$z$ color-color selection regions (effect \#1 above), low-$z$
interlopers with intrinsic colors {\it outside} the high-$z$ selection
criteria may get scattered into the selection regions by random
photometric errors.  The importance of such scatter could be crudely
gauged using simulations.  However, a more direct and robust approach
is to note that the photometric errors of \R\s27 objects in our
imaging are similar to those of \R\s25.5 objects in the Steidel et
al.\ surveys.  Because of this similarity we can expect expect similar
interloper fractions in our survey as in theirs.  The interloper
fractions measured spectroscopically by Steidel et al.\ (1999, 2003,
2004) include {\it both} the photometrically-scattered interlopers
being discussed here {\it and} the interlopers with intrinsic colors
that place them in the high-$z$ color-color selection regions (effect
\#1 above).  We can therefore conclude that the sum of {\it both} 
classes of interlopers in our survey amounts to $\lesssim$5\% at
\zs1.7, 2.2., and 3, and $\lesssim$ 20\% at \zs4.

\item {\it High-$z$ galaxies scattered out of the high-$z$ samples due to photometric errors.}
In addition to low-$z$ objects being scattered by photometric errors
into our color-color-selected samples (item \#2 above), true high-$z$
objects with intrinsic colors that should place them in these samples
may be scattered out of the slection regions because of random
photometric errors.  To first order such scatter should be no larger
than the scatter in the opposite direction (\#2 above), given that the
high-$z$ galaxies are less numerous at a given apparent magnitude than
low-$z$ ones.  However, the amount of such scatter can be gauged more
accurately using Monte Carlo simulations and we will use such
simulations as needed --- for example when we use these data to study
the high-$z$ galaxy luminosity functions (Sawicki \& Thompson 2004).

\item {\it High-$z$ galaxies with intrinsic colors that place them outside our color-color selection criteria.}  
Finally, there exist high-$z$ galaxies whose {\it intrinsic} colors
lie outside the color-color selection regions defined by
Equations~\ref{z4sel.eq}--\ref{z17sel.eq}.  For example, sufficient
amounts of interstellar dust will redden high-$z$ galaxies out of our
samples, moving them to the upper right in
Figures~\ref{GRIcolorcolor.fig} and \ref{UGRcolorcolor.fig}.  We have
no way here to directly assess the size of such a missed population.
We note, however, that our \UGRI\ selection ensures that our fainter
high-$z$ samples misse {\it exactly the same} classes of high-$z$
galaxies as are missed in the brighter work by the Steidel group.
Therefore --- unlike other optical LBG surveys that have to combine
bright with faint samples selected using different filter sets and
color-color selection criteria, we are free of {\it differential} bias
in our sample selection.  If we are biasing ourselves against certain
classes of objects, we are doing so in the same way as the Steidel et
al.\ samples, with no dependence on luminosity between our and their
work.

\end{enumerate}

Above we have discussed the ways in which objects may be scattered in
and out of our high-$z$ samples by photometric errors, and the ways in
which our samples may be systematically contaminated and biased.  We
are greatly aided in determining our interloper fractions by our
\UGRI\ selection that is analogous to the Steidel et al.\ work.  This 
identical selection is a key feature of our survey and confers upon us
a great advantage over other deep surveys that use different selection
criteria that have not been extensively tested and calibrated with
spectroscopy.

\section{SUMMARY AND DISCUSSION}\label{summarydiscussion}

In this paper we have introduced the Keck Deep Fields, a very deep
\UGRI\ imaging survey which we use to construct samples of very faint
star-forming galaxies at \zs4, \zs3, \zs2.2, and \zs1.7.  The key
features of this survey are:

\begin{enumerate}

\item The KDF survey uses the very same \UGRI\ filter set and 
spectroscopically-confirmed and -optimized color-color selection
techniques developed by Steidel et al. (1999, 2003, 2004), thus
obviating the need for expensive spectroscopic characterization of the
sample and allowing us to confidently select {\it faint} star-forming
galaxies at \zs4, \zs3, \zs2.2, and \zs1.7.

\item The completeness limit of the KDF is \Rlim\s27 (with small 
field-to-field variations), where \Rlim is the magnitude at which 50\%
of point sources are detected.  Because optically-selected high-$z$
galaxies are unresolved in our ground-based images, this magnitude
limit is also the 50\% completeness limit for high-$z$ galaxies in our
survey.

\item The KDF survey reaches up to \s1.5 magnitudes deeper than the 
wider-area, but shallower, imaging used by Steidel and collaborators,
allowing us to select samples of much fainter, substantially
sub-\Lstar\ objects at \zs4, \zs3, \zs2.2, \zs1.7 than are possible in
the Steidel et al.\ surveys.

\item To \R=27, the KDF survey contains 427, 1481, 2417, and 2043, 
\UGRI-selected star-forming galaxies at \zs4, \zs3, \zs2.2, and \zs1.7, 
respectively.

\item The KDF survey covers 169 \sqam\ and is split into three 
widely-separated, spatially-independent patches on the sky. It thereby
provides a large sample of high-$z$ star-forming galaxies whose
statistics are dominated neither by Poisson noise nor by cosmic
variance.

\end{enumerate}

Our survey complements directly the wider but shallower surveys by the
Steidel team by extending their well-understood slection techniques to
galaxies that are up to four times fainter than the limit of the
Steidel et al.\ work.  The depth and efficiency of the KDF stems from
three factors: (1) In obvious contrast to the Steidel et al.\ surveys,
which are typically carried out on 4m-class telescopes, our survey was
undertaken on a much larger, 10m-aperture telescope; to first order,
this simple increase in collecting area allows us to reach 6.25 times
deeper per unit exposure time making our survey practical.  (2)
Additionally, we used a two-channel instrument, that allowed us to
observe the same field in two wavebands simultaneously, thereby
greatly decreasing the amount of total telescope time required.  (3)
Finally, in our second observing run --- the run that yielded the bulk
of our data --- we benefited significantly from a very efficient,
UV-optimized detector mosaic, that significantly reduced the necessary
exposure times in \G-, and --- especially ---
\U-bands.

In a great many respects, the KDF survey is the best currently
available for the study of {\it faint} star-forming galaxies at
$z$\s2, 3 and 4. It holds very significant advantages over other very
deep imaging surveys.  For example, our survey is comparable in depth
to the FORS Deep Field (FDF; Heidt et al.\ 2003), but covers \s3.5
times more area and is distributed over three spatially-independent
patches compared to the FDF's single 48 \sqam\ field.  Our area is \s7
times smaller than the Subaru Deep Surveys 1200 \sqam\ (SDS; Ouchi et
al.\ 2004), but whereas the SXDF's $BVRi'$ filters allow the selection
of only $z$\s4 Lyman Break galaxies, our \UGRI\ filter set lets us
probe the \s2.2~Gyr time-span from $z$\s4 to \zs1.7; given our
redshift coverage, we are in a position to study not just the
properties but also the time evolution of the LBG population.  The HST
$UBVI$ imaging of the Hubble Deep Fields (HDFs; Williams et al.\ 1996,
Casertano et al.\ 2002) matches ours in wavelength coverage and
surpasses it in depth. However, the two HDFs combined cover a total of
only \s10 \sqam, or a mere \s6\% of the area of our survey.
Consequently, the HDFs are limited by Poisson noise in the number of
objects and are more susceptible to the effects of large-scale
structure.  The Hubble Ultra Deep Field (UDF), deeper still than the
HDFs, is similarly restricted to only a single small pointing, and,
moreover, lacks deep $U$-band coverage, meaning that it is restricted
to higher redshifts only.  In addition to our advantages of area
and/or wavelength coverage over {\it all} these surveys, one must also
add the extremely important advantage that our survey gains from its
use of the well-tested \UGRI\ filter system.  Our use of this filter
system and its associated color-color selectios allows us to
confidently select high-$z$ galaxies and tie them {\it directly} to
the work of the Steidel group at brighter fluxes.

To our knowledge, the only real competitor for the KDF survey is the
GOODS project.  The HST $BVIz$ imaging of the GOODS fields (Giavalisco
et al.\ 2004), which covers twice the area of our survey and to a
somewhat greater depth, provides an excellent dataset at
$z$$\gtrsim$4.  However, at \zs3 and below, the lack of uniform
$U$-band coverage of the GOODS fields makes them less useful since
only the northern GOODS field has been imaged in $U$-band to a depth
approaching that of our survey ($\sim$40 hrs in \s1.25$\arcsec$ seeing
with the KPNO 4m+Mosaic; Mauro Giavalisco, private communication).
Thus, the GOODS $U$-band imaging covers an area nearly identical to
that of our survey (160 \sqam\ vs.\ our 169 \sqam) to a similar depth.
However, this single GOODS field is potentially more affected by
cosmic variance than is the sum of our three spatially independent
patches.  Moreover, as always, our KDF \UGRI\ data holds the very
significant advantage at all redshifts \za 2--4 of being a {\it direct
and straightforward} extension of the spectroscopically-calibrated
Steidel et al.\ selection technique.  We therefore conclude that while
the GOODS HST data dominates the field above $z$=4, the KDF are better
suited for work at $z$$\lesssim$4.

In the present paper --- the first in a series --- we have introduced
our survey, described our observations and data reductions, and have
shown our selection criteria for high-$z$ star-forming galaxies.  As
we have argued above, in many ways ours is the best survey to study
the population of {\it faint} star-forming galaxies from \zs4 to
\zs1.7.  The key features of our survey are its combination of depth with 
the well-understood \UGR\ and \GRI\ color-color selection.  

With the survey introduced and the data described in the present
paper, subsequent papers in this series will address in detail the
properties and evolution of the population of very faint star-forming
galaxies as the Universe ages by 2.5$\times$ over the
\s2.2~Gyr from \zs4 to \zs1.7.


\vspace{5mm}

We dedicate this work to the memory of Bev Oke, one of whose great
many contributions to astronomy was the LRIS imaging spectrograph (Oke
et al.\ 1995) without which this work would not have happend.  We also
thank the Caltech time allocation committee for a generous time
allocation that made this project possible and the staff of the W.M.\
Keck Observatory for their help in obtaining these data.  We are also
grateful to Chuck Steidel for his encouragement and support of this
project and to Jerzy Sawicki for a thorough reading of the manuscript
and many useful comments.  Finally, we wish to recognize and
acknowledge the very significant cultural role and reverence that the
summit of Mauna Kea has always had within the indigenous Hawaiian
community; we are most fortunate to have the opportunity to conduct
observations from this mountain.

\newpage



\clearpage




\begin{deluxetable}{lccccc}
\tablecaption{\label{field-details.tab} Field details}
\tablehead{
\colhead{} &              
\colhead{02A} & 
\colhead{03A} &
\colhead{03B} &
\colhead{09A} &
\colhead{09B} \\
}
\startdata
RA (2000)\tablenotemark{a}  	&  02:09:41.3 &  03:21:38.6 &  03:21:57.6 &  09:33:36.9 &  09:33:35.8        \\
Dec (2000)\tablenotemark{a}	& $-$04:37:46 & $-$04:19:19 & $-$04:19:10 & $+$28:48:19 & $+$28:43:58         \\
size\tablenotemark{b}	   	& 7.00\arcmin$\times$4.93\arcmin & 7.15\arcmin$\times$5.26\arcmin & 6.75\arcmin$\times$4.81\arcmin & 6.89\arcmin$\times$5.07\arcmin & 6.92\arcmin$\times$5.24\arcmin \\ 
$E(B-V)$\tablenotemark{c}   	& 0.024 & 0.031 & 0.031 & 0.018 & 0.018 \\
$t_{exp}$(\U)\tablenotemark{d}	& 9600s \tablenotemark{e} & 21600s \tablenotemark{f} & 14400s \tablenotemark{e} & 22800s \tablenotemark{f} & 18000s \tablenotemark{e} \\
$t_{exp}$(\G)\tablenotemark{d}	& 7200s \tablenotemark{e} &  9600s \tablenotemark{e} & 10800s \tablenotemark{e} & 12000s \tablenotemark{f} & 10800s \tablenotemark{e} \\
$t_{exp}$(\R)\tablenotemark{d}	& 6300s \tablenotemark{g} & 21870s \tablenotemark{g} & 13125s \tablenotemark{g} & 25335s \tablenotemark{g} & 12075s \tablenotemark{g} \\
$t_{exp}$(\I)\tablenotemark{d}	& 3575s \tablenotemark{g} &  9720s \tablenotemark{g} & 10100s \tablenotemark{g} & 10440s \tablenotemark{g} &  6500s \tablenotemark{g} \\
seeing (\U)\tablenotemark{h}    & 0.98\arcsec & 1.38\arcsec & 0.93\arcsec & 1.03\arcsec & 0.97\arcsec \\
seeing (\G)\tablenotemark{h}    & 0.99\arcsec & 0.78\arcsec & 0.83\arcsec & 1.10\arcsec & 1.05\arcsec \\
seeing (\R)\tablenotemark{h}    & 0.80\arcsec & 1.09\arcsec & 0.80\arcsec & 0.82\arcsec & 0.82\arcsec \\
seeing (\I)\tablenotemark{h}    & 0.89\arcsec & 1.30\arcsec & 1.00\arcsec & 0.93\arcsec & 0.88\arcsec \\
common smoothed seeing\tablenotemark{i}& 1.0\arcsec  & 1.4\arcsec  & 1.0\arcsec  & 1.1\arcsec  & 1.1\arcsec \\
$\mu_{lim}$(\U)\tablenotemark{j}& 30.4 & 30.4 & 30.7 & 31.0  & 30.6 \\
$\mu_{lim}$(\G)\tablenotemark{j}& 30.5 & 30.6 & 31.0 & 30.8  & 30.8 \\
$\mu_{lim}$(\R)\tablenotemark{j}& 29.3 & 30.1 & 29.9 & 30.0  & 29.9 \\
$\mu_{lim}$(\I)\tablenotemark{j}& 28.7 & 29.3 & 29.3 & 29.5  & 29.5 \\
\R$_{lim}$\tablenotemark{k} 	& 26.66 & 26.76 & 27.21 & 27.31 & 27.14 \\
 N(22.5$\leq$\R$\leq$25.0)\tablenotemark{l}	& 30.3	& 29.0	& 30.1	& 22.6 	& 26.6 \\
\enddata
\tablenotetext{a}{Approximate coordinates of the field center}
\tablenotetext{b}{Includes regions of overlap between 03A and 03B, and 09A and 09B}
\tablenotetext{c}{Foreground dust attenuation in Schlegel et al.\ (1998) maps}
\tablenotetext{d}{Exposure time in the final, stacked image}
\tablenotetext{e}{Taken with the UV-optimized EEV CCD mosaic}
\tablenotetext{f}{Taken with the blue-side engineering-grade SITe CCD}
\tablenotetext{g}{Taken with the SITe/Tektronix 2k$\times$2k CCD}
\tablenotetext{h}{Median of a number of unsaturated point sources}
\tablenotetext{i}{Seeing in the final, stacked and smoothed images}
\tablenotetext{j}{1$\sigma$ surface brightness limit in mag/arcsec$^2$}
\tablenotetext{k}{Magnitude at which 50\% of point sources are detected in the stacked (but not smoothed) image}
\tablenotetext{l}{Number of objects per \sqam\ detected in the magnitude range \R=22.5--25.0}
\end{deluxetable}


\end{document}